\documentclass[12pt]{spieman}  
\usepackage{amsmath,amsfonts,amssymb}
\usepackage{graphicx}
\usepackage{setspace}
\usepackage{tocloft}

\title{Dirac, Schr\"odinger, and Maxwell equations in scalar and vector field quantum mechanics}

\author{Boris Chichkov}
\affil{Leibniz University Hannover, Institute of Quantum Optics, Welfengarten Srt. 1, 30167 Hannover, Germany }

\cftpagenumbersoff{figure}
\cftpagenumbersoff{table} 
\begin{document} 
\maketitle

\begin{abstract}
The quantum theory of relativistic particles, based on the first quantization technique similar to that used by Schr\"odinger and Dirac in formulating quantum mechanics, is reconsidered on the basis of a photon-like dispersion relation corresponding to the energy conservation equation of Einstein's special relativity. First, scalar quantum mechanics of particles operating with their wave functions is discussed. Using the first quantization of the photon-like dispersion relation, very simple new derivation of the Dirac equation is given.  Then, vector field quantum mechanics is introduced, which defines  vector fields associated with the relativistic particle.  Basic equations for the vector-field quantum mechanics, similar to the source-free Maxwell equations, are derived. Following these equations, the particle’s de Broglie wave can be considered as the transversal electromagnetic wave. Therefore, the "wave-particle duality" can be redefined as the ”electromagnetic wave-particle duality”.
Relationships between the scalar and vector field quantum mechanics are analyzed. 
\end{abstract}


{\noindent \footnotesize\textbf{*} Adress correspondence to \linkable{chichkov@iqo.uni-hannover.de} }


\section{Introduction}

Quantum non-relativistic  Schr\"odinger and relativistic Dirac equations are  among the most important scientific equations of the twentieth century. More old Maxwell equations derived in the nineteenth century have remained valid to this day and play a very important role  in their modern mathematical form introduced by Heaviside\cite{Hunt}. In this paper we demonstrate how all these equations can be derived using the first quantization technique of the photon-like dispersion relation corresponding to the energy conservation equation of Einstein's special relativity. First quantization deals with a fixed number of particles and converts classical energy conservation equations into quantum wave equations by replacing classical observables like energy and momentum by operators acting on the scalar wave function or vector fields. In this paper we provide a new and very simple derivation of the Dirac equation and its non-relativistic limits (Schr\"odinger and Levy-Leblond equations) representing standard quantum mechanics dealing with the operators acting on the scalar wavefunctions which  define probabilities to find a particle in a certain quantum state with a certain degree of localization in time and in space, in a volume $V>\lambda^3$, where $\lambda$ is the De Broglie wavelength of this particle.  Then, using the same dispersion relation and defining operators acting on vector functions,
we show how vector field quantum mechanics can be introduced for relativistic particles. We derive basic equations for the vector field quantum mechanics which are similar to source-free Maxwell equations.  After that we analyze the relationships between the scalar and vector field quantum mechanics. 

It should be noted that Hans Sallhofer, who was a student of Erwin Schr\"odinger, has derived the Dirac equation from calssical source-free Maxwell equations, written for a postulated medium with spatial dispersion, and established the so-called Maxwell-Dirac isomorphism\cite{Sall}. This isomorphism was confirmed in many other publications devoted to investigations of the links between the Maxwell and Dirac equations \cite{Sim1,Rash,Khol}, including the concepts of Optical Dirac equation\cite{Bar,Hor,Den,Min} and electronic Maxwell equations\cite{Ming,Iwo}.

A comprehensive discussion of different derivations of the Dirac equation can be found in the recent review.\cite{Sim} Solutions and symmetry  properties of the Dirac equation, including  the formalism of second quantization, can be found in many textbooks.\cite{Thal,Sch}

\section{Relativistic scalar quantum mechanics}
\subsection{Klein-Gordon and Dirac equations}

For a free relativistic particle, the particle energy $E$, its momentum ${\bf p}$, and its rest mass $m$ are connected by the following equation of Einstein's theory of special relativity
\begin{equation}\label{free}
E^2=({\bf p}c)^2+(mc^2)^2,  
\end{equation}
where $c$ is the speed of light. First quantization of this equation can be performed directly by replacing $E$ with the energy operator $ \hat E=i\hbar \partial/\partial t$ and  ${\bf p}$ with the momentum operator ${\bf \hat p}=-i\hbar\partial/\partial {\bf r}=-i\hbar \nabla $  and results in the Klein-Gordon equation
\begin{equation}\label{KG}
\left[\frac{\partial^2}{\partial t^2}-c^2\Delta +\frac{(mc^2)^2}{\hbar^2}\right]\Psi({\bf r},t)=0. 
\end{equation}
In case of a particle with $m=0$, the above equation reduces to the wave equation for a photon 
\begin{equation}\label{we}
\left(\frac{\partial^2}{\partial t^2}-\frac{c^2}{n^2}\Delta\right)\Psi({\bf r},t)=0,
\end{equation}
with the refractive index $n=1$. Note that for an ultra-relativistic particle with $E>>mc^2$,  Eq. (\ref{free}) can be replaced by the approximate relation $E\simeq pc$, which is similar to the dispersion relation for a photon in free space, and  Eq. (\ref{KG}) can be approximated by the wave equation.

For a relativistic particle moving in a potential field with the potential energy $V$, we have  
\begin{equation}\label{ec}
E=\sqrt{({\bf p}c)^2+(mc^2)^2} +V. 
\end{equation}
The corresponding dispersion relation between the photon energy  $E=\hbar\omega$ and its momentum ${\bf p}=\hbar {\bf k}$ in a dielectric medium 
with the refractive index $n=\sqrt{\epsilon\mu}$, where  $\epsilon$ and $\mu$ are the medium permittivity and permeability,   is given by   
\begin{equation}\label{ec1}
E=\frac{pc}{n}, \quad\quad 
p=|{\bf p}| =\frac{\sqrt{\epsilon\mu}E}{c}. 
\end{equation}
The formula on the right side is the Minkowski expression for the photon momentum which is widely used in photonics, laser physics, and nonlinear optics, and is in complete agreement with the special relativity.\cite{me}  
It is remarkable that for an arbitrary particle, Eq. (\ref{ec}) can be represented in the form of Eq. (\ref{ec1}) by introducing an effective medium with spatial dispersion. To define characteristics of this medium, we substitute  the particle momentum (\ref{ec1})  in Eq. (\ref{ec}) obtaining the following relation 
\begin{equation}\label{ec2}
(E-V)^2=\epsilon\mu E^2 + (mc^2)^2,  
\end{equation}
which allows to define the refractive index of the effective dispersive medium
\begin{equation}\label{n2}
n^2=\epsilon\mu=\frac{(E-V)^2-(mc^2)^2}{E^2}=\left( 1-\frac{V+mc^2}{E}\right)\left( 1-\frac{V-mc^2}{E}\right).  
\end{equation}
Now we have two formal possibilities for the definition of the effective medium permittivity and permeability.  We chose the following one
\begin{equation}\label{eps1}
\epsilon=\left( 1-\frac{V+mc^2}{E}\right), \quad\quad  \mu=\left( 1-\frac{V-mc^2}{E}\right).  
\end{equation}
The fact that the relativistic energy conservation Eq. (\ref{ec}) for arbitrary particle can be written in the form similar to the photon dispersion relation Eq. (\ref{ec1}) is very important and allows to get simple derivations of the known equations and to make interesting conclusions about the vector fields around this particle.  

We can re-write the photon-like dispersion relation Eq. (\ref{ec1}) between the energy and momentum of the relativistic particle  in two equivalent forms
\begin{equation}\label{ece}
\epsilon E=\sqrt{\frac{\epsilon}{\mu}}\;cp, \quad\quad 
\mu E=\sqrt{\frac{\mu}{\epsilon}}\;cp. 
\end{equation}
Multiplying the first equation by the function $\phi=\phi({\bf r},t)$ and the second one by the function $\chi=\chi({\bf r},t)$ and choosing the following relation between these functions $\chi =\sqrt {\epsilon/\mu}\;\phi$, Eqs. (\ref{ece}) can be replaced by
\begin{equation}\label{clas}
(E-V-mc^2) \phi =cp \chi, \quad\quad 
(E-V+mc^2) \chi =cp\phi,  
\end{equation}
where Eqs.(\ref{eps1}) for $\epsilon$ and $\mu$ were used.  Now introducing $p=({\bf e_k \cdot p)}$, where  ${\bf e_k}={\bf k}/k$ is the unit vector, and the energy and momentum operators, the above equations can be written in the Schr\"odinger form with the Hamiltonian operator $\hat H$
\begin{equation}\label{sch}
i\hbar\frac{\partial}{\partial t}\Psi ({\bf r}, t)=\hat H\Psi ({\bf r}, t), \quad \quad \Psi = \begin{pmatrix}
       \phi\\
       \chi
   \end{pmatrix},\quad \quad \hat H=\begin{pmatrix}
       V+mc^2 & c{\bf e_k \cdot \hat p}\\
        c{\bf e_k \cdot \hat p} & V-mc^2
   \end{pmatrix}.
\end{equation}
These quantum $2\times2$ matrix equations can be considered as the generalized Klein-Gordon equation for a spinless particle in the potential field, where similar to the Dirac interpretation the $\phi$ function determines particle and $\chi$ function anti-particle  probabilities. 

For characterization of particles having spin $s=1/2$ or $s=1$ it is convenient to introduce the Pauli matrices
\begin{equation}
   \sigma_x=\begin{pmatrix}
       0 & 1\\
       1& 0
   \end{pmatrix} ,\; \sigma_y=\begin{pmatrix}
       0 & -i\\
       i & 0
   \end{pmatrix}, \; \sigma_z=\begin{pmatrix}
       1 & 0\\
       0 & -1
   \end{pmatrix}
\end{equation}
and the Pauli vector $\boldsymbol{\sigma}=\sigma_x{\bf e}_x +\sigma_y{\bf e}_y+\sigma_z{\bf e}_z$, where ${\bf e}_{x,y,z}$ are the unit coordinate vectors. For us the following expressions for the scalar product of the Pauli vector with the particle momentum are important
\begin{equation}\label{sigp}
   \boldsymbol{\sigma}\cdot{\bf p}=\begin{pmatrix}
       p_z & p_x-ip_y\\
        p_x+ip_y& -p_z
   \end{pmatrix} ,\quad (\boldsymbol{\sigma}\cdot{\bf p})^2=\begin{pmatrix}
       {\bf p}^2 & 0\\
        0 & {\bf p}^2 
   \end{pmatrix} = {\bf p}^2 I,
\end{equation}
where $I$ is the $2\times2$ identity matrix. 
Now using the dispersion relation (\ref{ec1}) written in the quadratic form and multiplied by the identity matrix $I$, we get   
\begin{equation}
\epsilon\mu E^2 I= c^2 {\bf p}^2 I = c^2 (\boldsymbol{\sigma}\cdot{\bf p})^2.  
\end{equation}
This equation can be viewed as a product of two identical matrix equations, similar to Eqs. (\ref{ece})
\begin{equation}\label{ecem}
\epsilon E I=\sqrt{\frac{\epsilon}{\mu}}\;c(\boldsymbol{\sigma}\cdot{\bf p}), \quad\quad 
\mu E I=\sqrt{\frac{\mu}{\epsilon}}\;c(\boldsymbol{\sigma}\cdot{\bf p}). 
\end{equation}
Multiplying the first equation by the spinor $\Phi=\Phi({\bf r},t)=\begin{pmatrix}
       \phi_1\\
        \phi_2
   \end{pmatrix}$ and the second one by the spinor $X =X({\bf r},t)=\begin{pmatrix}
       \chi_1\\
        \chi_2
   \end{pmatrix}$ and choosing the following relation between them $X =\sqrt {\epsilon/\mu}\;\Phi$, Eqs. (\ref{ecem}) can be replaced by
\begin{equation}\label{clasD}
(E-V-mc^2) \Phi =c(\boldsymbol{\sigma}\cdot{\bf p}) X, \quad\quad 
(E-V+mc^2) X =c(\boldsymbol{\sigma}\cdot{\bf p})\Phi,  
\end{equation}
Introducing the energy and momentum operators, the above equations become the Dirac equations for the $\Phi$ and $X$ spinors
\begin{eqnarray}\label{Dirac1}
i\hbar\frac{\partial}{\partial t}\Phi =(V+mc^2)\Phi +c(\boldsymbol{\sigma}\cdot{\bf \hat p}) X, \\\quad\quad 
i\hbar\frac{\partial}{\partial t}X =(V-mc^2)X +c(\boldsymbol{\sigma}\cdot{\bf \hat p}) \Phi.  
\end{eqnarray}
Introducing the 4-component wave function $\Psi=\Psi({\bf r},t)=\begin{pmatrix}
       \Phi\\
        X
   \end{pmatrix}$, one can write the Dirac equation in the $4\times 4$ matrix form
\begin{equation}\label{schm}
i\hbar\frac{\partial}{\partial t}\Psi ({\bf r}, t)=\hat H\Psi ({\bf r}, t), \quad \quad 
   \hat H=\begin{pmatrix}
       I(V+mc^2) & c (\boldsymbol{\sigma}\cdot {\bf \hat p}) \\
       c (\boldsymbol{\sigma}\cdot {\bf \hat p}) & I(V-mc^2)
   \end{pmatrix}.
\end{equation}  
To the best of my knowledge, this new derivation of the Dirac equation, using the photon like dispersion relation Eq. (\ref{ec1}), is one of the most simple,  straightforward, and physically justified (see the recent review of different derivations of the Dirac equation\cite{Sim}). 
The Dirac equation can also be written with respect to the particle kinetic energy $E_k=E-mc^2$, instead of the full energy $E$, using the following transformation $\Psi=\Psi_k \exp (-imc^2t/\hbar)$
\begin{equation}\label{schm}
i\hbar\frac{\partial}{\partial t}\Psi_k ({\bf r}, t)=\hat H_k\Psi_k ({\bf r}, t), \quad \quad 
   \hat H_k=\begin{pmatrix}
       IV & c (\boldsymbol{\sigma}\cdot {\bf \hat p}) \\
       c (\boldsymbol{\sigma}\cdot {\bf \hat p}) & I(V-2mc^2)
   \end{pmatrix}.
\end{equation}

\subsection{Non-relativistic limits, Schr\"odinger and Levy-Leblond equations}

The well-known non-relativistic limit of Eq. (\ref{ec}) is defined by 
\begin{equation}\label{ecnr}
E\simeq mc^2 +\frac{\bf{p}^2}{2m}+V=mc^2+E_{cl}. 
\end{equation}
This allows to replace Eqs. (\ref{clas}) by
\begin{equation}\label{clas1}
(E_{cl}-V) \phi =cp \chi, \quad\quad 
2mc^2 \chi =cp\phi.  
\end{equation}
By inserting $\chi=  p \phi/(2mc) $ in the left equation and $\phi= cp/(E_{cl}-V)) \chi$ in the right equation, we get the following equation for both $\phi$ and $\chi$ functions 
\begin{equation}\label{clas2}
(E_{cl}-V) \phi =\frac{{\bf p}^2}{2m}\phi,  
\end{equation}
which after the replacement of $E_{cl}$ with the operator $\hat E_{cl}=ih\partial/\partial t$ and $\bf p$ with the standard momentum operator results in the Schr\"odinger equation. Thus, in the non-relativistic limit both $\phi$ and $\chi$ functions satisfy the Schr\"odinger equation.

In the non-relativistic limit, Eqs.(\ref{clasD}) for spinors  will be replaced by 
\begin{equation}\label{clasD1}
(E_{cl}-V) I \Phi =c(\boldsymbol{\sigma}\cdot{\bf p}) X, \quad\quad 
2mc^2 I X =c(\boldsymbol{\sigma}\cdot{\bf p})\Phi,  
\end{equation}
From the second equation one can see that $X=(\boldsymbol{\sigma}\cdot{\bf p})\Phi/(2mc)$. That is why $\Phi$ are called large Dirac components and $X$ small.
After the introduction of operators, the above equations become the Lévy-Leblond equations \cite{Levy} for a non-relativistic particle with spin, which represent the non-relativistic limit of the Dirac equations. For both $\Phi$ and $X$ spinors, Eqs. (\ref{clasD1}) can be written in the following form 
\begin{equation}\label{clasD2}
(E_{cl}-V)I \Phi =\frac{({\boldsymbol{\sigma}\cdot\bf p})^2}{2m}\Phi.  
\end{equation}
Introducing operators and taking into account Eq. (\ref{sigp}), one can see that in the non-relativistic limit 
all spinor components and all components of the Dirac wave function satisfy the Schr\"odinger equation.

\section{Vector field quantum mechanics}

In this section we discuss how the vector field quantum mechanics can be introduced on the basis of first quantization of Eq. (\ref{ec1}), $E=pc/n$, with the refractive index determined by Eq. (\ref{n2}). In vector field quantum mechanics, we introduce operators acting on vector fields, which are associated with a relativistic particle.  For the particle energy, the operator remains the same, $ \hat E=i\hbar \partial/\partial t$, and we need to define an operator for the absolute value of the particle momentum $p=|{\bf p}|=\sqrt{{\bf p}^2}$.  We consider a class of vector functions perpendicular to the particle momentum, $\boldsymbol{\Psi}({\bf r},t)\perp {\bf p}$, with the scalar product  ${\bf p}\cdot\boldsymbol{\Psi}=0$, corresponding to $\nabla \boldsymbol{\Psi}=\rm{div}\boldsymbol{\Psi}=0$ in the operator form.  For this class of functions,  the operator $\hat p$ satisfying the following condition $\hat p^2\boldsymbol{\Psi}={\bf \hat p}^2\boldsymbol{\Psi}$ can be defined as
$\hat p=\hbar \nabla\times=\hbar{\rm {rot}}=i{\bf \hat p}\times$, where ${\bf \hat p}$ is the standard momentum operator. One can see that the above condition is satisfied
\begin{equation}
\hat p^2\boldsymbol{\Psi}=\hbar^2\rm {rot\,(rot}\boldsymbol{\Psi})=\hbar^2[\rm {grad\,(div}\boldsymbol{\Psi})-\Delta\boldsymbol{\Psi}]=-\hbar^2\Delta \boldsymbol{\Psi}={\bf \hat p}^2\boldsymbol{\Psi}.
\end{equation}

By multiplying Eqs. (\ref{ec1}) by the $\boldsymbol{\Psi}$ function and introducing the energy and momentum $\hat p$ operators, we can derive the Schr\"odinger equation in the vector form for a relativistic particle
\begin{equation}\label{VecS}
i\hbar\frac{\partial}{\partial t}\boldsymbol{\Psi}({\bf r}, t)=\hat H\boldsymbol{\Psi}({\bf r}, t), \quad\quad \hat H=\frac{c}{n}\hbar \nabla\times=i\frac{c}{n}{\bf \hat p}\times
\end{equation}
with the Hamiltonian operator acting on the vector field. This equation formally coincides with the vector Schr\"odinger equation for a photon which has been used before for the derivation of Maxwell equations.\cite{Chichkov2}  A similar procedure can be used here. To simplify derivations, we make a step back and replace in Eq. (\ref{VecS}) the energy and momentum $\bf \hat p$ operators by the energy $E$ and momentum $\bf p$, respectively
\begin{equation}\label{Vecback}
E\boldsymbol{\Psi}({\bf r}, t)=i\frac{c}{n}{\bf  p}\times\boldsymbol{\Psi}({\bf r}, t). 
\end{equation}
The $\boldsymbol{\Psi}$ vector function can be represented in the following form $\boldsymbol{\Psi}=\sqrt{\epsilon}{\bf E}+i\sqrt{\mu}{\bf H}$, where ${\bf E}$ and ${\bf H}$ are the transversal vector fields. In this form  $\boldsymbol{\Psi}$ is known as the Riemann–Silberstein vector or Weber 
vector\cite{Bi, Seb}. ${\bf E}$ and ${\bf H}$  belong to the class of vector fields satisfying the following conditions:  ${\bf p}\cdot {\bf E}={\bf p}\cdot {\bf H}=0$ and  $\rm{div}{\bf {E}}=\rm{div}{\bf {H}}=0$. These fields are similar to the transversal electric and magnetic fields.  Separating the real and imaginary parts in Eq. (\ref{Vecback}) and taking into account $n=\sqrt{\epsilon\mu}$, we get the following equations
\begin{equation}\label{Maxwell}
\epsilon E{\bf E} =-c{\bf p}\times {\bf H} , \quad\quad 
\mu E {\bf H} = c{\bf  p}\times {\bf E},  
\end{equation}
where $E$ is the particle energy.  Introducing Eqs, (\ref{eps1}) for $\epsilon$ and $\mu$, we get 
\begin{equation}\label{Maxwell1}
(E-V-mc^2){\bf E} =-c{\bf p}\times {\bf H} , \quad\quad 
(E-V+mc^2) {\bf H} = c{\bf  p}\times {\bf E}.  
\end{equation}
Using the operators for energy and momentum, we arrive at the quantum version of Maxwell equations for the oscillating fields around an arbitrary particle
\begin{eqnarray}
i\hbar\frac{\partial}{\partial t}{\bf E} -(V+mc^2){\bf E} &=& (i\hbar\, c)\;{\rm  rot}\, {\bf H}, \quad\quad\;\; {\rm div\, {\bf E}=0 } \label{Maxwell2}\\
i\hbar\frac{\partial}{\partial t}{\bf H} -(V-mc^2) {\bf H} &=& -(i\hbar\, c)\;{\rm  rot}\, {\bf E}, \quad\quad  {\rm div\, {\bf H}=0 }.\label{Maxwell3}
\end{eqnarray}
These are the basic equations for the vector-field quantum mechanics. 
In case of $m=0$ and $V=0$ these equations reduce to the ordinary Maxwell equations (in the CGS Gaussian system of units) for a photon in free space.\cite{Chichkov2} For photons representing the entire spectrum of electromagnetic radiation, the photon - electromagnetic wave duality is the well-known and well-established fact. The above equations suggest that this should be also valid for arbitrary quantum particles. The particle’s  de Broglie wave can be considered as the transversal electromagnetic wave corresponding to the "electromagnetic wave-particle duality". So far, the existence of such electromagnetic de Broglie waves has not been experimentally confirmed.

It has been discussed earlier by Hans Sallhofer\cite{Sall} that from classical Maxwell equations written in the form of Eqs. (\ref{Maxwell2}, \ref{Maxwell3}) the Dirac equation can be derived. This can be done by scalar multiplication of the above equations on the Pauli vector \cite{Sall}. This procedure will be illustrated in the next Section.

For an electron in Coulomb potential $V=-Ze^2/r$, where $Z$ is the nuclear charge, Eqs.   (\ref{Maxwell2}, \ref{Maxwell3}) describe oscillating electric and magnetic fields in a hydrogen atom ($Z=1$) or hydrogen ion. In this case the hydrogen atom can be considered as a spherical resonator, defined by the Coulomb potential, with the trapped standing electromagnetic wave. The structure of the transversal electromagnetic vector fields in the hydrogen atom and its spectrum have also been considered by Hans Sallhofer.\cite{Sall1} 

In the non-relativistic limit, using Eq. (\ref{ecnr}), we can re-write Eqs. (\ref{Maxwell1}) in the following form
\begin{equation}\label{Maxwell4}
(E_{cl}-V) {\bf E} =-c{\bf p}\times {\bf H} , \quad\quad 
(2mc^2) {\bf H} = c{\bf  p}\times {\bf E}.  
\end{equation}
Taking into account ${\bf p}\cdot {\bf E}={\bf p}\cdot {\bf H}=0$ and the following relation
${\bf  p}\times[{\bf  p}\times {\bf E}]=-{\bf  p}^2{\bf E}$, which is also valid for ${\bf H}$, we can present Eqs.(\ref{Maxwell4})  in the equivalent form
\begin{equation}\label{Maxwell5}
(E_{cl}-V) {\bf E} =\frac{\bf p^2}{2m}{\bf E} , \quad\quad 
(E_{cl}-V) {\bf H} =\frac{\bf p^2}{2m}{\bf H}.  
\end{equation}
Introducing the momentum and energy operators, we obtain the following mixed quantum Schr\"odinger-Maxwell equations for the fields in the non-relativistic limit 
\begin{eqnarray}
i\hbar\frac{\partial}{\partial t}{\bf E} &= &{\Big[}-\frac{\hbar^2}{2m}\,\nabla^2+V{\Big ]} {\bf E}, \quad\quad\ {\rm div\, {\bf E}=0 } \label{Maxwell6}\\
i\hbar\frac{\partial}{\partial t}{\bf H}& =& {\Big [}-\frac{\hbar^2}{2m}\,\nabla^2+V{\Big ]} {\bf H}, \quad\quad  {\rm div\, {\bf H}=0 }.\label{Maxwell7}
\end{eqnarray}

\section{Relationships between the scalar and vector field quantum mechanics}

To derive relations between the scalar and vector field quantum mechanics, we return to Eqs.(\ref{Maxwell1}) and perform scalar multiplication of these equations with the Pauli vector 

\begin{equation}\label{MaxwellP}
(E-V-mc^2)\;(\boldsymbol{\sigma}\cdot{\bf E}) =-c\boldsymbol{\sigma}\cdot({\bf p}\times {\bf H}) , \quad\quad 
(E-V+mc^2)\;(\boldsymbol{\sigma}\cdot {\bf H}) = c\;\boldsymbol{\sigma}\cdot({\bf  p}\times {\bf E}).  
\end{equation}
Using the following relation between the Pauli vector and two arbitrary orthogonal vectors $\bf a$ and $\bf b$, satisfying ${\bf a}\cdot{\bf b}=0$
\begin{equation}
i\boldsymbol{\sigma}\cdot({\bf a}\times {\bf b})=(\boldsymbol{\sigma}\cdot {\bf a})(\boldsymbol{\sigma}\cdot {\bf b})
\end{equation}
and applying it to $\bf p, E$ and $\bf p, H$ vector couples, we get
\begin{equation}\label{MaxwellP1}
(E-V-mc^2)\;(\boldsymbol{\sigma}\cdot{\bf E}) =ic(\boldsymbol{\sigma}\cdot{\bf p})(\boldsymbol{\sigma}\cdot{\bf H}) , \quad\quad 
(E-V+mc^2)\;(\boldsymbol{\sigma}\cdot {\bf H}) = -ic(\boldsymbol{\sigma}\cdot{\bf p})(\boldsymbol{\sigma}\cdot{\bf E}).  
\end{equation}
By comparing these equations with Eqs. (\ref{clasD}) for spinors $\Phi$ and $X$, one can see that they are equivalent when
\begin{equation}\label{clasDirac}
\Phi \sim (\boldsymbol{\sigma}\cdot{\bf E}), \quad\quad 
 X \sim i(\boldsymbol{\sigma}\cdot{\bf H}).  
\end{equation}
According to these equations, we have two possibilities for the relation between the 4 components of the Dirac wave function and the corresponding transversal vector fields
\begin{equation}\label{DM}
\Psi=\begin{pmatrix}
       \Psi_1 \\
       \Psi_2 \\
       \Psi_3\\
       \Psi_4
   \end{pmatrix}=\begin{pmatrix}
       \phi_1 \\
       \phi_2 \\
       \chi_1\\
       \chi_2
   \end{pmatrix}\sim \begin{pmatrix}
       E_z \\
       E_x+iE_y \\
       iH_z\\
       iH_x-H_y
   \end{pmatrix}, \quad \quad
\begin{pmatrix}
       \phi_1 \\
       \phi_2 \\
       \chi_1\\
       \chi_2
   \end{pmatrix}\sim \begin{pmatrix}
       E_x-iE_y \\
       -E_z \\
       iH_x+H_y\\
       -iH_z
   \end{pmatrix}.
\end{equation}  
This formal isomorphism between the Dirac and Maxwell equations has been previously noted by Hans Sallhofer\cite{Sall} and confirmed by other authors\cite{Sim1,Rash,Khol,Bar,Hor,Den,Min} without explicit derivations of the vector field quantum Maxwell equations.

It can be seen that the following transformation of fields 
\begin{equation}
{\bf E}\rightarrow i\sqrt{\frac{\mu}{\epsilon}}{\bf H},\quad \quad   {\bf H}\rightarrow -i\sqrt{\frac{\epsilon}{\mu}}{\bf E},
\end{equation}
is equivalent to the transformation of Dirac spinors $\Phi\rightarrow\sqrt{\mu/\epsilon} X$ and $X\rightarrow \sqrt{\epsilon/\mu}\Phi$.

\section{Conclusions}
Scalar and vector field quantum mechanics of relativistic particles based on the first quantization technique of the photon-like dispersion relation, corresponding to the energy conservation equation of Einstein’s special relativity, have been reconsidered. Using the same dispersion relation, the Dirac, Schr\"odinger, and Maxwell equations have been derived. Vector field quantum mechanics, presented by the source-free Maxwell equations, predicts the existence of transversal electromagnetic fields associated with quantum particles. This provides a new physical interpretation of the wave function in terms of the corresponding electromagnetic field. Therefore, the particle’s de Broglie wave can be considered as the transversal electromagnetic wave, which allows to redefine the "wave-particle duality"  as the ”electromagnetic wave-particle duality”. This view simplifies our understanding of the fact that  photons (bosons), electrons (fermions), and other massive particles produce the same interference patterns in the double slit experiments. 

\subsection* {Acknowledgments}
I acknowledge financial support from the  DFG, German 
Research Foundation under Germany’s Excellence Strategy within
the Cluster of Excellence PhoenixD (EXC 2122, Project ID
390833453) and the Cluster of Excellence QuantumFrontiers
(EXC 2123, Project ID 390837967).  I thank Dr. A.  Evlyukhin for critical discussions.



\begin{thebibliography}{99}
\bibitem{Hunt} B.J. Hunt, Oliver Heaviside: A first-rate oddity, {\it Physics Today} {\bf 65}, 48-54 (2012).
\bibitem{Sall} H. Sallhofer, Elementary Derivation of the Dirac Equation. X, {\it Z. Naturforsch.} A {\bf 41}, 468-470 (1986).
\bibitem{Sim1} V.M. Simulik, I.Y. Krivsky, Relationship between maxwell and dirac equations: symmetries, quantization, models of atom, {\it Rep. Math. Phys.} {\bf 50}(3), 315-328 (2002).
\bibitem{Rash} SA Rashkovskiy, Classical-field model of the hydrogen atom, {\it Indian J. Phys.} {\bf 91}, 607–621 (2017).
\bibitem{Khol} A.L. Kholodenko, Maxwell-Dirac Isomorphism Revisited: From Foundations
of Quantum Mechanics to Geometrodynamics and Cosmology, {\it Universe} {\bf 9}, 288 (2023).
\bibitem{Bar} S.M. Barnett, Optical Dirac equation, {\it New J. Phys.} {\bf 16}, 093008 (2014).
\bibitem{Hor} S.A.R. Horsley, Topology and the optical Dirac equation, {\it Phys. Rev. A}, {\bf 98}, 043837 (2018).
\bibitem{Den} M.R. Dennis, T. Tijssen, and M.A. Morgan, On the Majorana representation of the optical
Dirac equation, {\it J. Phys. A: Math. Theor.} {\bf 56}, 024004 (2023).
\bibitem{Min} M. Li and S.A.R Horsley, The electronic and electromagnetic Dirac equations, {\it New J. Phys.} {\bf 26}, 023007 (2024).
\bibitem{Ming} M. Li, P. Shi, L. Du, and X. Yuan, Electronic Maxwell’s equations, {\it New J. Phys.} {\bf 22},  113019 (2020).
\bibitem{Iwo} I. Bialynicki-Birula, Comment on ‘Electronic Maxwell’s equations’, {\it New J. Phys.}, {\bf 22}, 113019 (2020).
\bibitem{Sim} V.M. Simulik, The Dirac equation near centenary: a contemporary introduction to the Dirac
equation consideration, {\it J. Phys. A: Math. Theor.} {\bf 58},  053001 (2025).
\bibitem{Thal} B. Thaller, The Dirac Equation, (Springer, Berlin, 1992).
\bibitem{Sch} F. Schwable, Advanced Quantum Mechanics, 3rd edn. (Springer, Berlin, 2000).
\bibitem{me} N.B. Chichkov and B.N. Chichkov, On the origin of photon mass, momentum, and
energy in a dielectric medium [Invited], {\it Optical Material Express} {\bf 11}, 2722 (2021).
\bibitem{Levy} 
J-M. Lévy-Leblond, Nonrelativistic Particles and Wave Equations, {\it Commun. math. Phys.} {\bf 6}, 286--311 (1967).
\bibitem{Chichkov2} B. N. Chichkov, On the first quantization and quantum diversity of photons, {\it Advanced Photonics} {\bf 7}, 055001 (2025).
\bibitem{Bi} I. Bialynicki-Birula and Z. Bialynicki-Birula, The role of the Riemann–Silberstein vector in classical and quantum theories of electromagnetism, {\it J. Phys. A: Math. Theor.} {\bf 46}, 053001 (2013).
\bibitem{Seb} C.T. Sebens, Electromagnetism as Quantum Physics, {\it Foundations of Physics} {\bf 49}, 365 (2019).
\bibitem{Sall1} H. Sallhofer, Hydrogen in Electrodynamics. VI The General Solution {\it Z. Naturforsch.} A {\bf 45}, 1361-1366 (1990).
\end{thebibliography}
\end{document}